\documentclass[twocolumn,showpacs,preprintnumbers,amsmath,amssymb,aps,prl]{revtex4-1}

\usepackage{graphicx}
\usepackage{dcolumn}
\usepackage{bm}
\usepackage{amsmath,amssymb}
\usepackage{epstopdf}
\usepackage[T1]{fontenc}
\usepackage{hyperref}
\usepackage{tabularx}
\newcommand\numberthis{\addtocounter{equation}{1}\tag{\theequation}}

\newcolumntype{Y}{>{\centering\arraybackslash}X}
\begin{document}

\preprint{}

\title{Acoustic-Soft Optical Mode Coupling Leads to Negative Thermal Expansion of GeTe near the Ferroelectric Phase Transition}
\author{{\DJ}or{\dj}e Dangi{\'c}\textsuperscript{1,2}}
\email{djordje.dangic@tyndall.ie}
\author{Aoife R. Murphy\textsuperscript{1,2}}
\author{\'Eamonn D. Murray\textsuperscript{3}}
\author{Stephen Fahy\textsuperscript{1,2}}
\author{Ivana Savi\'c\textsuperscript{2}}
\email{ivana.savic@tyndall.ie}
\affiliation{\textsuperscript{\normalfont{1}}Department of Physics, University College Cork, College Road, Cork, Ireland}
\affiliation{\textsuperscript{\normalfont{2}}Tyndall National Institute, Dyke Parade, Cork, Ireland}
\affiliation{\textsuperscript{\normalfont{3}}Department of Physics and Department of Materials, Imperial College London, London SW7 2AZ, UK}

\date{\today}

\begin{abstract}

GeTe is a well-known ferroelectric and thermoelectric material that undergoes a structural phase transition from a rhombohedral to the rocksalt structure at $\sim 600-700$ K. We present a first principles approach to calculate the thermal expansion of GeTe in the rhombohedral phase up to the Curie temperature. We find the minimum of the Helmholtz free energy with respect to the structural parameters, including the internal atomic displacement, in a manner similar to the traditional Gr{\"u}neisen theory, explicitly accounting for the variation of the static elastic energy with respect to all structural parameters. We obtain the temperature variation of the structural parameters of rhombohedral GeTe in very good agreement with experiments. In particular, we correctly reproduce a negative volumetric thermal expansion of GeTe near the phase transition. We show that the negative thermal expansion is induced by the coupling between acoustic and soft transverse optical phonons, which is also responsible for the low lattice thermal conductivity of GeTe. 

\end{abstract}


\maketitle

\section{I. Introduction} 

Most materials expand upon heating, while those that shrink are much less common. Recent interest in these materials with negative thermal expansion (NTE) is also driven by technological applications that require materials with zero thermal expansion across a desired temperature range \cite{nte_review_1,nte_review_2,nte_review_3}. Even though NTE is an unusual phenomenon, it is relatively common for materials near structural phase transitions, and is typically associated with soft phonons and strong anharmonicity \cite{nte_review_1,nte_review_2,nte_review_3}. 

The Gr{\"u}neisen theory \cite{Bryce,Keblinski,PhysRevB.71.205214,thermexpSnSe,PhysRevB.94.054307,Arash_NTE} is the standard approach to calculate thermal expansion from first principles, using density functional theory. In this method, anharmonicity of the crystal potential is described via mode Gr{\"u}neisen parameters (GP's), which represent the changes of phonon frequencies with volume \citep{Bryce,Keblinski,PhysRevB.71.205214,thermexpSnSe,PhysRevB.94.054307,Arash_NTE}. Negative GP's of certain phonon modes are commonly identified as the source of NTE \cite{NTE1,NTE2, NTE3, NTE3, Arash_NTE}. Phonon frequencies and mode GP's are usually calculated using the harmonic approximation. First principles methods that describe phonon frequency renormalization due to anharmonicity have been recently developed, such as the self consistent harmonic approximation (SCHA) {\citep{SCPA}} and temperature dependent effective potentials (TDEP) \cite{STDEP}. These and related approaches were recently used to describe the negative thermal expansion of ScF$_{3}$~\cite{Ambroaz} and Si~\cite{PNAS_Hellman_2018}. In principle, these methods are capable of modeling thermal expansion of materials near phase transitions. However, to the best of our knowledge, no previous work has investigated this possibility.

\begin{figure}[h]
\begin{center}
\includegraphics[width = 0.49\textwidth]{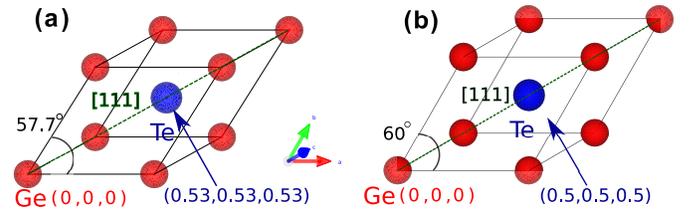}
\end{center}
\caption{Primitive unit cell of GeTe at (a) 0 K and (b) above the Curie temperature, generated using \textsc{VESTA} software \cite{VESTA}. The low temperature rhombohedral structure becomes more similar to the rocksalt structure as temperature increases: the angle between the primitive lattice vectors $\theta$ becomes closer to 60$^{\circ}$ and the Te internal atomic position $(\tau,\tau,\tau)$ approaches $(0.5,0.5,0.5)$ in reduced coordinates.}
\label{fig1}
\end{figure}

GeTe is the simplest ferroelectric material that exhibits NTE near the phase transition \cite{main, newmain, Marchenkov1994, abrikosov}. This makes it an ideal test case for identifying the physical effects leading to NTE. At low temperatures, GeTe crystallizes in a rhombohedral structure \cite{main,newmain,Goldack}, characterized by the Te internal atomic displacement along the [111] direction from its high symmetry position in the rocksalt phase, (0.5,0.5,0.5) in reduced coordinates, see Fig.~\ref{fig1}. The angle between the primitive lattice vectors of the rhombohedral structure also differs from 60$^{\circ}$ for the rocksalt phase. GeTe experiences a structural phase transition from a rhombohedral to the rocksalt structure at $\sim 600-700$ K depending on the carrier concentration \cite{abrikosov}. This phase transition is mediated by softening of the zone center transverse optical (TO) mode \cite{STEIGMEIER19701275,Wdowik}, which corresponds to the frozen-in Te internal atomic displacement along the [111] axis.

The proximity to the ferroelectric phase transition also makes GeTe a very good thermoelectric material, either in the pure \cite{levin,gete-jacs,yaniv-gete-jap-16,biswas-gete-rev,natureasia-gete,pei-joule-gete,ZT_PT_PNAS_2018} or alloyed form \cite{GeSbte1,GeSbte2,GeSbte3, TAGS1,TAGS2,TAGS3, TAGS4, leadalloy1, ronanleadalloy}. Its soft TO modes interact strongly with acoustic modes which carry most heat, thus leading to the low lattice thermal conductivity \cite{ronanleadalloy} and the high thermoelectric figure of merit. The same mechanism is responsible for the exceptionally low lattice thermal conductivity of PbTe \cite{ssc148-417,nmat10-614,prb85-155203,ronanscatt}. GeTe can be driven closer to the soft TO mode phase transition not only by changing the temperature but also by alloying with PbTe \cite{pbgete}. We have recently shown that the acoustic-TO coupling is strongest for those (Pb,Ge)Te alloy compositions that are very near the phase transition, and leads to the minimal lattice thermal conductivity when mass disorder is neglected \cite{ronanleadalloy}.  

In this paper, we present a first principles method to compute the thermal expansion of the rhombohedral phase of GeTe up to the Curie temperature. We calculate the structural parameters by minimizing the total free energy with respect to each structural parameter in the spirit of the Gr{\"u}neisen theory. We explicitly include internal atomic position as an independent variable in the minimization process. Although this effect was included to some extent in previous calculations of thermal expansion \cite{Bryce,PhysRevB.71.205214,thermexpSnSe,PhysRevB.94.054307,Arash_NTE} by relaxing atomic positions due to applied strain, this may not be sufficient for materials near phase transitions. Our approach enables us to determine the temperature dependence of the static elastic energy variations with structural parameters, which we find is the key to correctly describing the thermal expansion of GeTe near the phase transition. We show that our calculated thermal evolution of the structural parameters of GeTe agrees well with experiments. Negative volumetric thermal expansion of GeTe near the phase transition is also well described in our model. We find that the coupling between acoustic and soft TO modes is the dominant mechanism leading not only to the low lattice thermal conductivity of GeTe, as shown previously, but also to its NTE. 
 

\section{II. Method}

We model the thermal expansion of rhombohedral GeTe using the ideas of the Gr{\"u}neisen theory within the elastic and harmonic approximations for the mechanical and vibrational properties of solids, respectively. A rhombohedral unit cell is defined with the primitive lattice vectors $a(b,0,c)$, $a(-\frac{b}{2},\frac{b\sqrt{3}}{2},c)$ and $a(-\frac{b}{2},-\frac{b\sqrt{3}}{2},c)$. Here $a$ is the lattice constant, and $b$ and $c$ are defined as:
\begin{align*}
b &= \sqrt{\frac{2}{3}(1-\cos\theta)}, \\
c &= \sqrt{\frac{1}{3}(1+2\cos\theta)}, \numberthis
\end{align*}
where $\theta$ is the angle between the primitive lattice vectors. The reduced atomic positions of GeTe within this unit cell are: ($0,0,0$) for Ge atom and ($\tau ,\tau ,\tau $) for Te atom. The temperature dependence of these structural parameters is implicit. The Helmholtz total free energy of a rhombohedral crystal per unit cell is defined as \cite{srivastava}: 
\begin{align*}
&F(a,\theta,\tau,T) = E_{el}(a,\theta,\tau) + F_{vib}(a,\theta,\tau,T), \numberthis
\end{align*}
where $E_{el}(a,\theta,\tau)$ and $F_{vib}(a,\theta,\tau,T)$ correspond to the static elastic and vibrational free energy at temperature $T$, respectively. The values of all the structural parameters at a certain temperature can be found by minimizing the total free energy with respect to each structural parameter $u$, $u\in (a,\theta,\tau)$: 
\begin{align*}
&\frac{\partial F}{\partial u} =\frac{\partial E_{el}}{\partial u} + \frac{\partial F_{vib}}{\partial u} = 0. \numberthis
\label{eq2}
\end{align*}

Within the harmonic approximation, vibrational free energy is given as \cite{srivastava}: 
\begin{align*}
&F_{vib}=\sum_{\textbf{q},s}\left[ \frac{\hbar \omega _{s}(\textbf{q})}{2} + k_{B}T\text{ln}\left(1-\exp\left(-\frac{\hbar \omega _{s}(\textbf{q})}{k_{B}T}\right)\right)\right],\numberthis
\end{align*}
where $\omega _{s}(\textbf{q})$ is the phonon frequency of mode $s$ and wave vector $\textbf{q}$, and $k_B$ is the Boltzmann constant. The derivative of vibrational free energy with respect to one of the structural parameters $u$ reads: 
\begin{align*}
\frac{\partial F_{vib}}{\partial u} =& -\frac{1}{u}\sum_{\textbf{q},s} \hbar \omega _{s}(\textbf{q}) \left(n(\omega _{s}(\textbf{q})) + \frac{1}{2}\right)\gamma ^u _{s} (\textbf{q}), \numberthis
\label{eq4}
\end{align*}
where $n(\omega _{s}(\textbf{q}))$ is the Bose-Einstein occupation factor at temperature $T$ for a phonon with frequency $\omega _{s}(\textbf{q})$. We define the generalized Gr{\"u}neisen parameters with respect to each structural parameter as:
\begin{align*}
\gamma ^u _{s}(\textbf{q}) = -\frac{u}{\omega _{s}(\textbf{q})}\frac{\partial \omega _{s}(\textbf{q})}{\partial u}. \numberthis
\label{eq5}
\end{align*}
We note that the generalized GP's $\gamma ^{u}_{s} (\mathbf{q})$ are computed without the relaxation of atomic positions with applied strain, in contrast to previous GP calculations~\cite{Bryce,Keblinski,PhysRevB.71.205214,thermexpSnSe,PhysRevB.94.054307}. This difference in the GP's definitions will lead to differences between our calculated GP values and those of prior work for GeTe~\cite{Bernuskoni}. Nonetheless, we account for the atomic relaxation effects via the generalized GP's $\gamma ^\tau _{s}(\textbf{q})$. This separation of variables allows us to explicitly track the coupling between the soft TO mode and strain, as we will show.

Phonon frequencies and generalized Gr{\"u}neisen parameters can be computed either using the harmonic approximation, or accounting for the phonon frequency renormalization due to anharmonicity and the temperature variation of structural parameters. Here we calculate phonon frequencies and generalized GP's for the values of the structural parameters $a$, $\theta$ and $\tau$ at $0$ K. This is a reasonable approximation since only the soft TO modes close to the zone center will have a considerable temperature dependence in GeTe. We expect that the temperature induced renormalization of soft TO modes will have a substantial effect on thermal expansion only very close to the phase transition.

The static elastic part of total free energy can be expanded in a Taylor series as: 
\begin{align*}
E_{el} = E_{0} + \sum_{u} K_{u}\Delta u + \sum_{u,v} K_{uv} \Delta u\Delta v. \numberthis
\label{eq6}
\end{align*}
$\Delta u$ and $\Delta v$ represent the small deviations of the structural parameters $u$ and $v$ from their equilibrium values for temperature $T$ ($u,v \in \{a,\theta,\tau\}$, $u \geq v$). We define the first and second order coefficients as the changes of $E_{el}$ with respect to the changes of structural parameters: $K_{u} = \frac{\partial E_{el}}{\partial u}$ and $K_{uv} =(1-\frac{1}{2} \delta _{uv} )\frac{\partial ^2 E_{el}}{\partial u \partial v}$. The relationship between these coefficients and elastic constants is discussed in Appendix A. The final form for the derivative of static elastic energy with respect to one of the structural parameters reads:
\begin{align*}
\frac{\partial E_{el}}{\partial u} =& K_{u} + \sum_{v} (1+\delta_{vu})K_{vu}\Delta v. \numberthis
\label{eq7}
\end{align*}

Coefficients $K$ change with temperature due to the contribution of the higher order terms in the Taylor expansion of static elastic energy. If we label the changes of the structural parameters at temperature $T$ with respect to their values at 0 K as:
\begin{align*}
\centering
\delta a &= a - a_{0}, \\
\delta \theta &= \theta - \theta _{0}, \numberthis \\
\delta \tau &= \tau - \tau _{0},
\end{align*} 
we can expand static elastic energy as:
\begin{align*}
\label{eq8}
&E_{el} = \sum_{u,v} K^{0}_{uv} (\Delta u + \delta u)(\Delta v + \delta v) + \\
& \sum_{u,v,w} K^{0}_{uvw} (\Delta u + \delta u)(\Delta v + \delta v)(\Delta w + \delta w) \numberthis + \\
& \sum_{u,v,w,t} K^{0}_{uvwt} (\Delta u + \delta u)(\Delta v + \delta v)\times\\&(\Delta w + \delta w)(\Delta t + \delta t).
\end{align*}
$K^{0} _{uv}$, $K^{0} _{uvw}$ and $K^{0} _{uvwt}$ are the second, third and fourth order coefficients defined for the changes of structural parameters calculated at 0 K, and $\delta u\in \{\delta a,\delta \theta,\delta \tau\}$ ($u \ge v \ge w \ge t$). From Eqs.~(\ref{eq6}) and (\ref{eq8}), we obtain coefficients $K_{u}$ and $K_{uv}$ that depend on the changes $\delta u$ from the $0$ K values, e.g.:
\begin{align*}
\label{eq9}
& K_{a}= 2K^{0} _{aa}\delta a + K^{0} _{a\tau}\delta \tau + K^{0} _{a\theta}\delta \theta + 3K^{0} _{aaa}\delta a^2 + \\
& 2(K^{0} _{aa\tau}\delta \tau + K^{0} _{aa\theta}\delta \theta)\delta a + K^{0} _{a\tau\tau}\delta \tau ^2 +\\
&K^{0} _{a\theta\theta}\delta \theta^2 + K^{0} _{a\theta\tau}\delta\theta\delta\tau + \text{terms with 4th order}~K^0,  \numberthis \\ 
&K_{aa} = K^{0} _{aa} + 3K^{0} _{aaa}\delta a + K^{0} _{aa\theta}\delta \theta + K^{0} _{aa\tau} \delta \tau +\\
&6K^{0} _{aaaa}\delta a^2  + 3(K^{0} _{aaa\theta}\delta \theta + K^{0} _{aaa\tau} \delta \tau)\delta a + K^{0} _{aa\theta\theta}\delta \theta ^2 +  \\ 
&K^{0} _{aa\tau\tau}\delta \tau ^2 + K^{0} _{aa\theta\tau}\delta \theta \delta \tau. 
\end{align*}

The temperature dependence of elastic coefficients $K_{uv}$ described by Eq.~\eqref{eq9} is directly related to the strength of anharmonic interactions involving very long wavelength acoustic and optical phonons. We thus effectively capture the anharmonic coupling between different zone center phonon modes up to the second order, including that between acoustic and soft transverse optical modes. Anharmonicity of the generalized GP's is taken into account only in the lowest order. We will show that this treatment of anharmonic effects is sufficient to describe the NTE of GeTe near the phase transition.

Substituting Eqs.~(\ref{eq4}) and (\ref{eq7}) into Eq.~(\ref{eq2}), we obtain:
{\small  
  \setlength{\abovedisplayskip}{6pt}
  \setlength{\belowdisplayskip}{\abovedisplayskip}
  \setlength{\abovedisplayshortskip}{0pt}
  \setlength{\belowdisplayshortskip}{3pt}
\begin{align*}
\Delta u = \sum_{v} S_{vu} \left[ \sum_{\textbf{q},s}\hbar \omega _{s}(\textbf{q})\left( n(\omega _{s}(\textbf{q})) + \frac{1}{2}\right)\frac{\gamma^{v} _{s}(\textbf{q})}{v} - K_{v}\right]. \numberthis
\label{eq10}
\end{align*}
}
$S_{vu}$ are the elements of the matrix defined as an inverse of the matrix of coefficients $\hat{K}$:
\begin{equation}
\hat{K} =
\begin{bmatrix}
2K_{aa} & K_{a\theta} & K_{a\tau} \\
K_{a\theta} & 2K_{\theta\theta} & K_{\theta\tau}  \\
K_{a\tau} & K_{\theta\tau} & 2K_{\tau\tau}
\end{bmatrix}.
\end{equation}
The matrix $\hat{S}$ is related to the compliance matrix which represents an inverse of the elastic constants matrix. We note that coefficients $K_u$ and $K_{uv}$ are functions of the structural parameters changes, $\delta a$, $\delta \theta$ and $\delta \tau$, see Eq.~(\ref{eq9}). We solve Eq.~(\ref{eq10}) for $\delta a$, $\delta \theta$ and $\delta \tau$ at each temperature by requiring that $\Delta u = 0$, which gives the thermal equilibrium structure. To do this, we construct an iterative solution as $\delta u_{i+1}=\delta u_i+\Delta u_i(\delta a_i, \delta \theta_i, \delta \tau_i)$, where $\Delta u_i$ is given by Eq.~(\ref{eq10}). This is iterated until $\Delta u_i\approx 0$.

We note that the presented method to calculate thermal expansion is inexpensive and straightforward to implement. Its implementation requires: (i) the density functional theory (DFT) calculations of the phonon frequencies and generalized Gr{\"u}neisen parameters for the $0$ K values of the structural parameters, (ii) the calculation of the DFT energy surface for a range of structural parameter values, whose fitting gives coefficients $K^0$ (Eq.~(\ref{eq8})), and (iii) the iterative solution for $\delta a$, $\delta \theta$ and $\delta \tau$ in Eq.~(\ref{eq10}) until $\Delta a$, $\Delta \theta$ and $\Delta \tau$ become zero for a range of temperatures. 

Our approach for obtaining the thermal expansion of rhombohedral materials near soft optical mode phase transitions can be linked to the standard method based on the Gr{\"u}neisen theory \cite{Bryce,Keblinski}, as shown in Appendix A. The standard approach finds the minimum of the total free energy of the system with respect to strain, rather than structural parameters. It includes the influence of atomic positions on total free energy by accounting for their relaxation due to applied strain. Far from the phase transition, our method fully corresponds to the standard one. However, the standard approach does not track the temperature dependence of internal atomic position and the corresponding static elastic energy changes, which are important for the accurate description of thermal expansion near the phase transition. More details about these differences can be found in Appendix A. On the other hand, establishing the precise relationship between our method and statistical mechanics approaches \cite{Rabe,Rabe2,FerroelecFunct} is less straightforward and requires further study.

\section{III. Computational details}

DFT calculations were performed using the plane wave basis set, the generalized gradient approximation with Perdew-Burke-Ernzerhof \cite{GGAPBE} parametrization  (\textsc{GGA-PBE}) for the exchange-correlation potential and the Hartwigsen-Goedecker-Hutter (HGH) pseudopotentials \cite{HGHpseudo} as implemented in \textsc{ABINIT} code \cite{ABINIT}. For the ground state and static elastic energy calculations, we used a 32 Hartree energy cutoff for plane waves and a four shifted $12\times 12\times 12$ \textbf{k}-point grid for Brillouin zone sampling of electronic states. Harmonic interatomic force constants at zero temperature were calculated from Hellmann-Feynman forces obtained by the finite difference supercell approach using \textsc{PHONOPY} code \cite{phonopy}. Forces were computed using 128-atom supercells ($4\times4\times4$ rhombohedral unit cells) with a 24 Hartree cutoff and a four shifted $3\times 3 \times 3$ \textbf{k}-point grid. Phonon frequencies were calculated using a $20\times 20\times 20$ \textbf{q}-point grid for vibrational modes. We obtained generalized Gr{\"u}neisen parameters using a finite difference method, taking the finite displacement to be smaller than 1\% for $a$, and smaller than 1\% of the difference between the 0 K rhombohedral and high temperature rocksalt structures for $\theta$ and $\tau$. For the calculation of coefficients $K^0$ in Eq.~(\ref{eq8}), we parametrized the energy surface on uniform grids for the values of structural parameters $a$, $\theta$ and $\tau$ from the 0 K rhombohedral structure to the high temperature rocksalt structure.
 

\section{IV. Results and discussion}

We calculated the structural parameters of GeTe at 0 K using \textsc{DFT} and two different exchange-correlation functionals, local density approximation (\textsc{LDA}) \cite{HGHpseudo} and \textsc{GGA-PBE}, see Table \ref{tb1}. To our knowledge, the measured values of the structural parameters at zero temperature are not available. Nevertheless, it is likely that, as the temperature is reduced from 295 K to 0 K, the angle and internal atomic position would deviate further from the high-symmetry (rocksalt) values, and would agree better with the GGA-PBE calculation than with the LDA. Since our goal is to describe the temperature dependence of structural parameters near the phase transition, where internal atomic position plays a crucial role, we use the \textsc{GGA-PBE} functional in all further calculations. Our values of structural parameters are also in good agreement with previous DFT calculations \cite{PhysRevB.95.024311, Wdowik}.

\begin{table}[h]
\begin{center}
\begin{tabularx}{0.5\textwidth}{ c | Y | Y | Y | Y }
\hline \hline
  & $a$ [\r{A}] &$\theta$ [deg]&$\tau$&$V_{0}$ [\r{A}$ ^{3}$] \\
 \hline
  \textsc{LDA}      & 4.207 & 58.788  & 0.524 & 51.193 \\ 
 \hline 
  \textsc{GGA-PBE}  & 4.381 & 57.776  & 0.530 & 56.420   \\
 \hline
 Experiment (295 K) & 4.299 & 57.931  & 0.525 & 53.513  \\
 \hline \hline
\end{tabularx}
\end{center}
\caption{Lattice parameters of GeTe at 0 K, calculated using \textsc{LDA} and \textsc{GGA-PBE} functionals, and compared with experimental results \cite{main}. $a$ stands for lattice constant, $\theta$ for angle, and $\tau$ for internal atomic coordinate.} 
\label{tb1}
\end{table}  

\begin{figure}
\begin{minipage}{0.48\textwidth}
\begin{center}
\includegraphics[width = 0.9\textwidth]{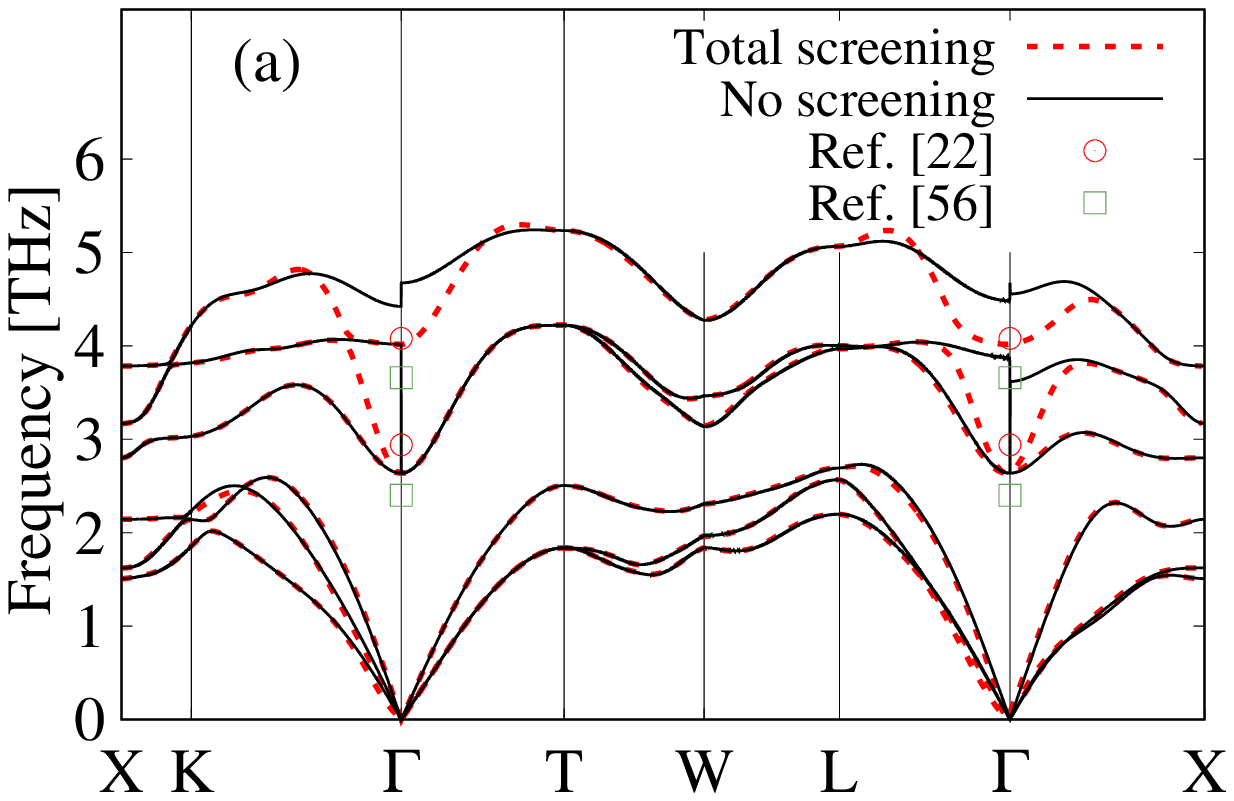}
\end{center}
\end{minipage}
\begin{minipage}{0.48\textwidth}
\begin{center}
\includegraphics[width = 0.9\textwidth]{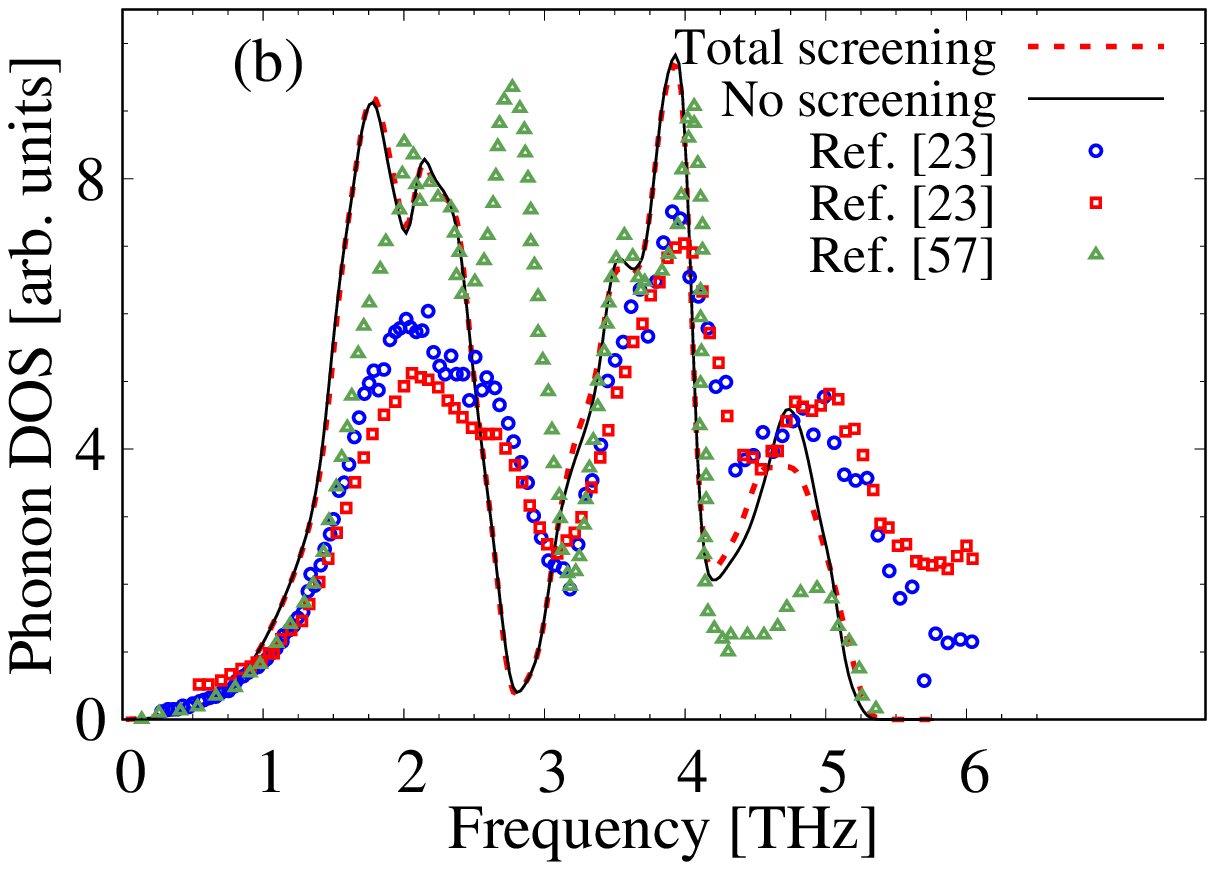}
\end{center}
\end{minipage}
\caption{(a) Phonon dispersion of GeTe calculated using \textsc{GGA-PBE} exchange-correlation functional neglecting and accounting for screening (solid black lines and dashed red lines, respectively). The frequencies of the zone centre Raman active modes were taken from the measurements of Ref.~\cite{STEIGMEIER19701275} (red circles) and Ref.~\cite{Fons} (green squares). (b) Phonon density of states of GeTe calculated neglecting and including screening (solid black line and dashed red line, respectively) and measured by Ref.~\cite{Wdowik} (blue circles and red squares) and Ref.~\cite{Pereira} (green triangles). The integral of the density of states over frequency is normalized to unity.}
\label{fig2}
\end{figure}

The phonon dispersion of GeTe at 0 K is given in Fig. \ref{fig2}(a), together with the experimental results for the frequencies of the Raman active zone center modes \cite{STEIGMEIER19701275, Fons}. Large intrinsic concentrations of charge carriers in real GeTe samples (1- 20$\times$10$^{20}$ cm$^{-1}$\cite{natureasia-gete}) completely screen long range interactions \cite{STEIGMEIER19701275}. We roughly estimate this effect by setting Born effective charges to zero in the calculation of phonon frequencies (see dashed red lines in Fig.~\ref{fig2}(a)). To evaluate the importance of screening, we also neglect this effect in the phonon calculation by using Born effective charge values obtained using density functional perturbation theory (DFPT) (solid black lines in Fig.~\ref{fig2}(a)). Using both approaches, our calculated phonon frequencies at the zone center agree very well with experimental results \cite{STEIGMEIER19701275, Fons}. Fig.~\ref{fig2}(b) illustrates that our computed phonon densities of states (DOS) of GeTe at 0 K compare fairly well with experiments \cite{Wdowik, Pereira}.  Since there are no appreciable differences in the calculated phonon DOS if we exclude or roughly include screening effects, we neglect screening in all further calculations \footnote{We verified explicitly that our treatment of screening produces a very small effect on the values of structural parameters with respect to the unscreened case. We expect that a more sophisticated treatment of screening will change these values more substantially, as observed experimentally in GeTe samples with different carrier concentrations~\cite{Marchenkov1994, abrikosov}.}. Our phonon dispersions of GeTe also agree well with a previous DFPT calculation \cite{PhysRevB.95.024311}.

\begin{figure}[ht!]
\begin{minipage}{0.49\textwidth}
\begin{center}
\includegraphics[width = 0.9\textwidth]{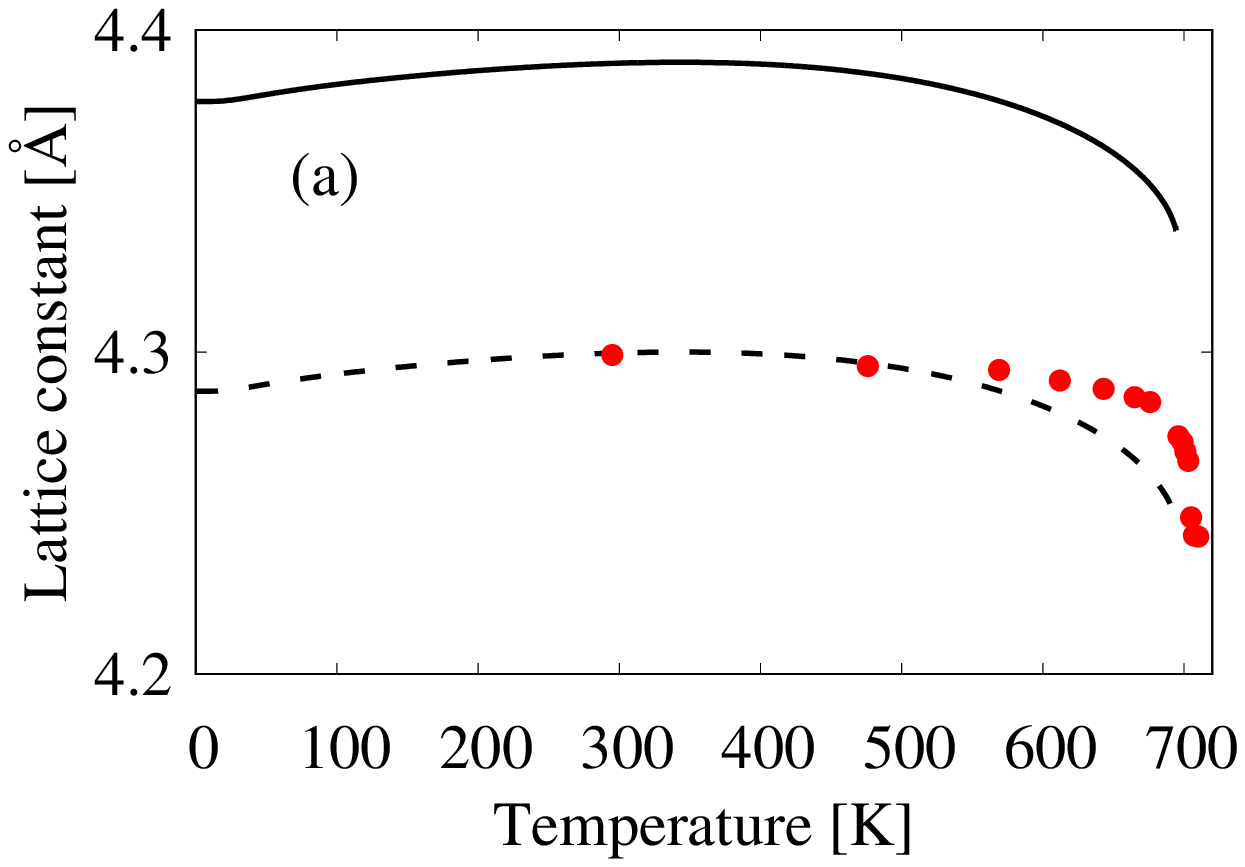}
\end{center}
\end{minipage}
\begin{minipage}{0.49\textwidth}
\begin{center}
\includegraphics[width = 0.9\textwidth]{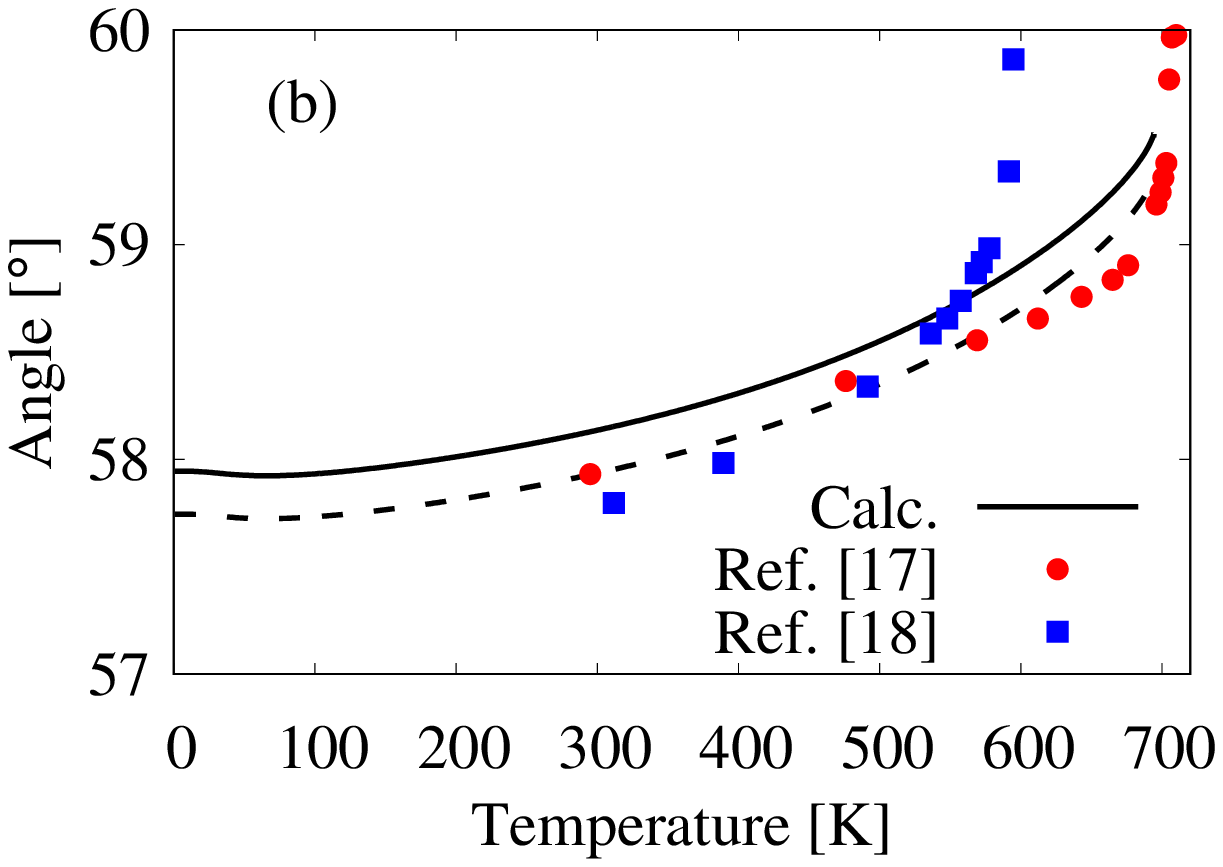}
\end{center}
\end{minipage}
\begin{minipage}{0.49\textwidth}
\begin{center}
\includegraphics[width = 0.9\textwidth]{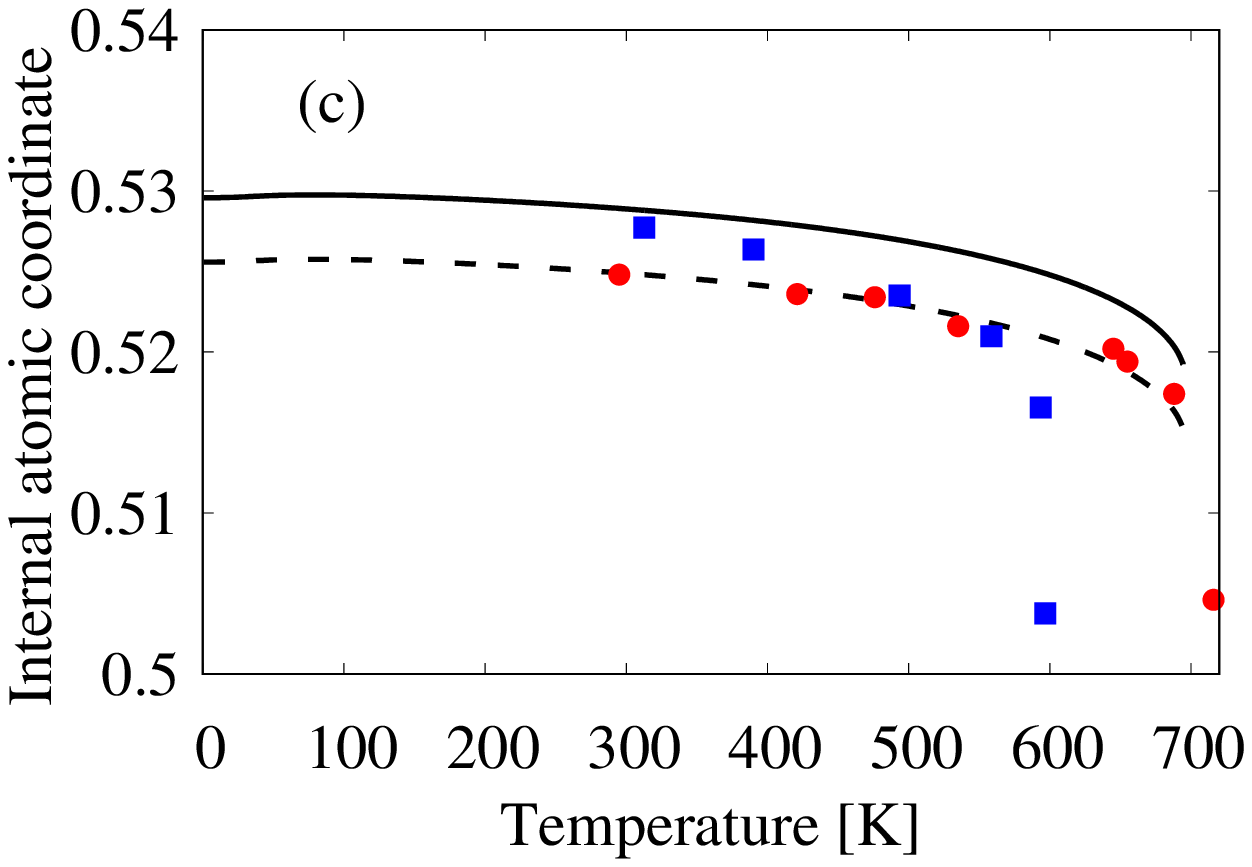}
\end{center}
\end{minipage}
\caption{Structural parameters of GeTe as a function of temperature: (a) lattice constant $a$, (b) angle $\theta$, and (c) internal atomic coordinate $\tau$. Solid black lines represent our calculations. Red circles and blue squares correspond to the measurements of Refs.~\cite{main} and \cite{newmain}, respectively. Dashed black lines represent our calculations shifted by the difference between our calculated values and the experimental values of Ref.~\cite{main} at 300 K.}
\label{fig3}
\end{figure}

The temperature dependence of all structural parameters of rhombohedral GeTe (lattice constant, angle and internal atomic coordinate $\tau$) are illustrated in Fig. \ref{fig3}. Solid lines represent our calculations, while symbols show the measurements of Refs.~\cite{main,newmain}. The experimental values were transformed from the pseudocubic to the rhombohedral unit cell for comparison with our results. The computed temperature variation of structural parameters is in good agreement with experiments, despite the small discrepancy between the \textsc{GGA-PBE} and the room temperature experimental structural parameters (see Table \ref{tb1}). Dashed lines in Fig.~\ref{fig3} represent our calculations shifted by the difference between our values and the experimental values of Ref.~\cite{main} at 300 K. The calculated temperature dependence of the zone center TO mode frequency (see Appendix B) is also in very good agreement with experiment \cite{STEIGMEIER19701275}. We highlight that all these agreements are obtained fully from first principles, without any empirical parameters.

Our calculated structural parameters of rhombohedral GeTe show clear indications of the ferroelectric phase transition near 700 K, see Fig. \ref{fig3}. As temperature increases, the angle $\theta$ and the internal atomic coordinate $\tau$ tend to their high symmetry values, 60$^0$ and 0.5, respectively. Moreover, the temperature dependence of all structural parameters diverges from a linear behavior at high temperatures (500-700 K), which signals the proximity to the phase transition. 

\begin{figure}[ht!]
\begin{minipage}{0.49\textwidth}
\begin{center}
\includegraphics[width = 0.9\textwidth]{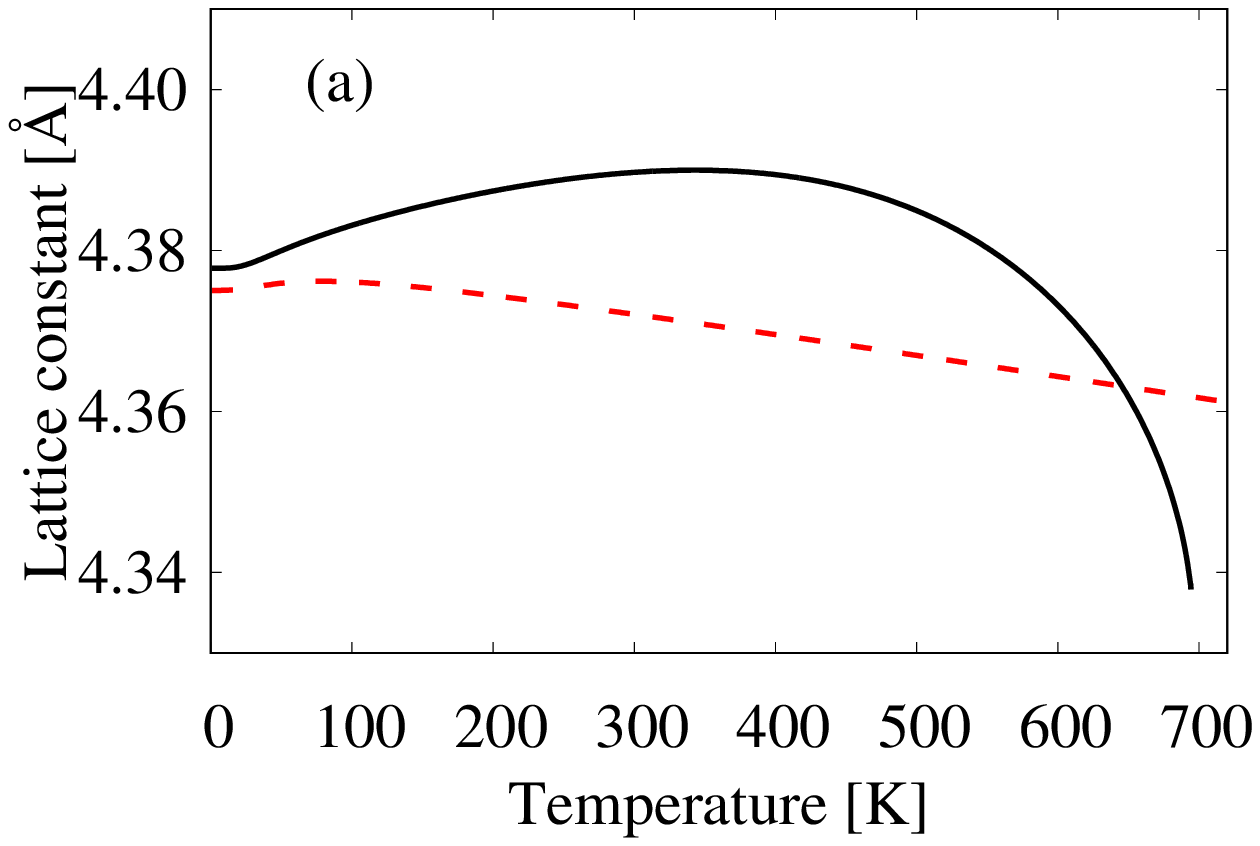}
\end{center}
\end{minipage}
\begin{minipage}{0.49\textwidth}
\begin{center}
\includegraphics[width = 0.9\textwidth]{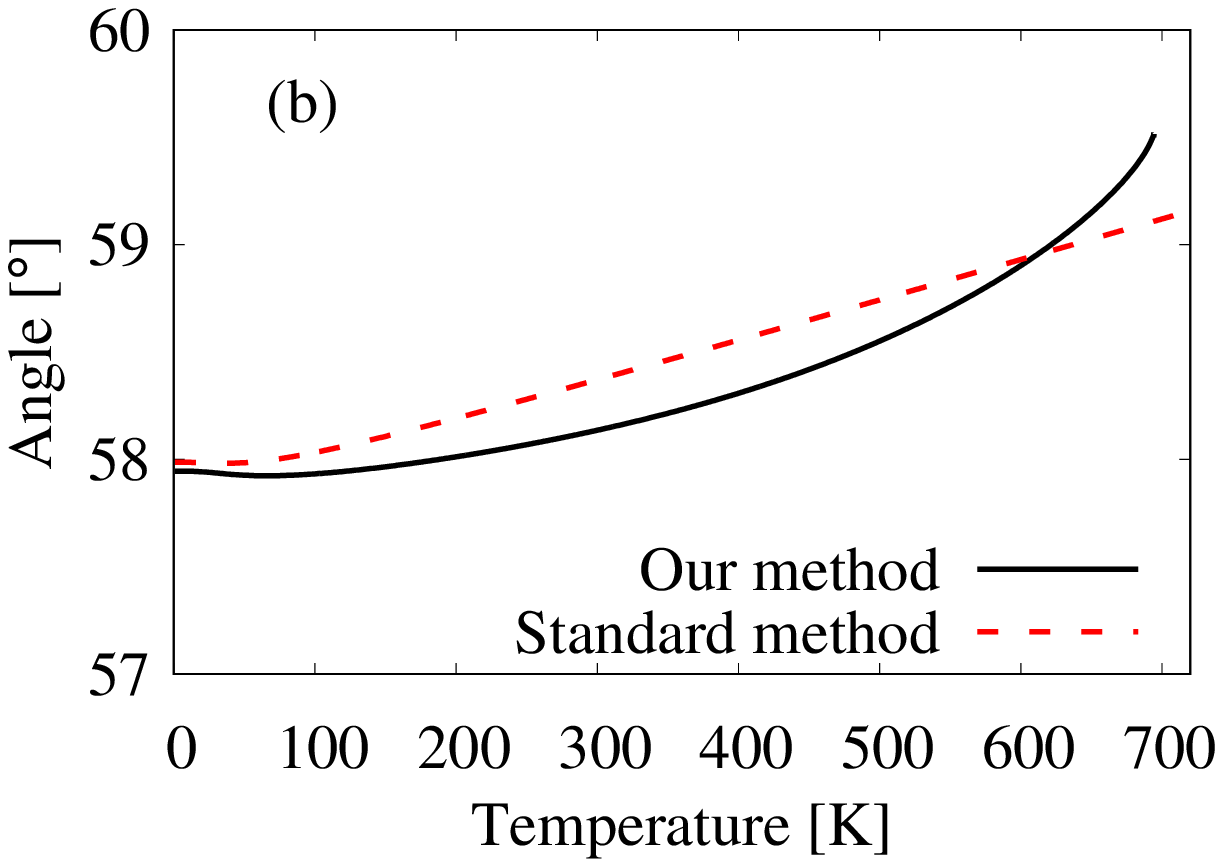}
\end{center}
\end{minipage}
\begin{minipage}{0.49\textwidth}
\begin{center}
\includegraphics[width = 0.9\textwidth]{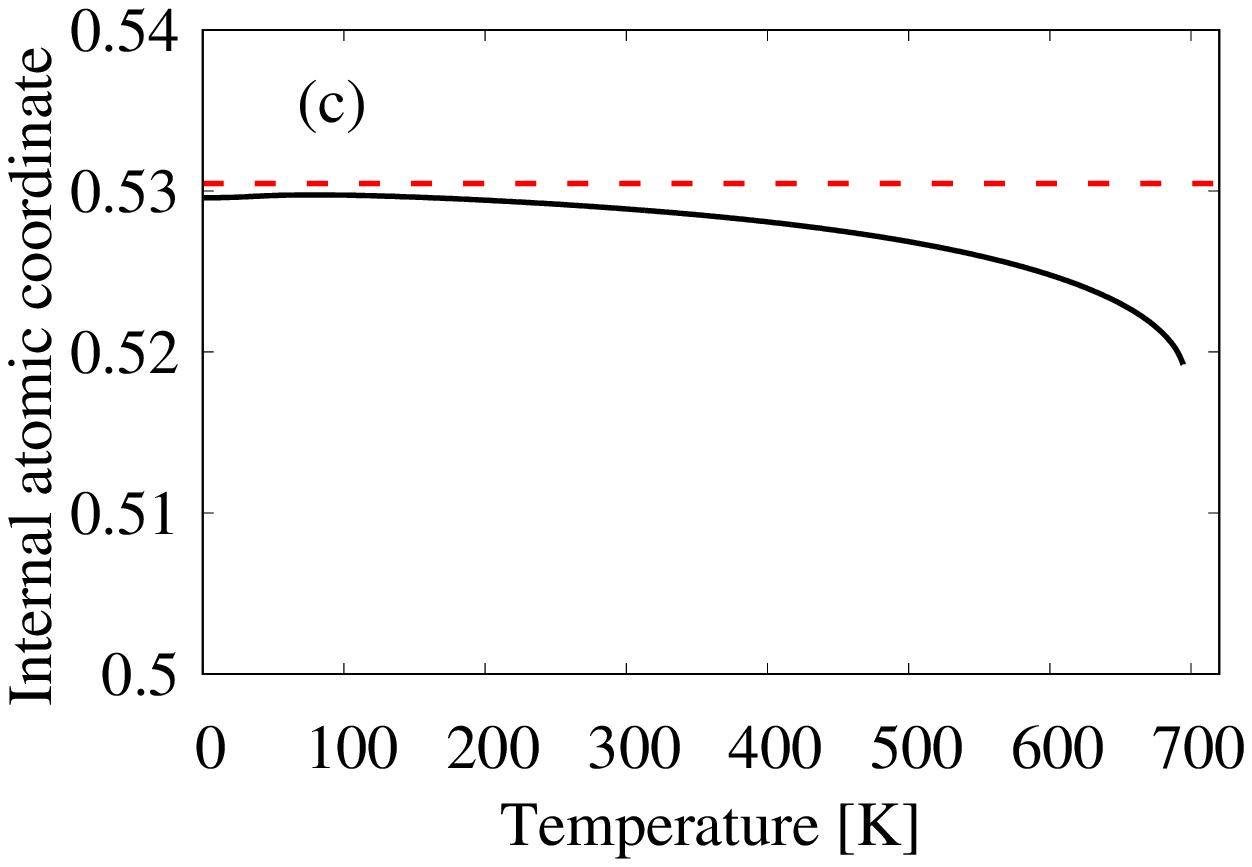}
\end{center}
\end{minipage}
\caption{Structural parameters of GeTe as a function of temperature: (a) lattice constant, (b) angle, and (c) internal atomic coordinate. Solid black lines represent the results obtained using our approach, while dashed red lines correspond to the standard method (see text for full explanation).}
\label{fig4}
\end{figure}

The thermal evolution of the structural parameters of GeTe is correctly captured only when the total free energy is minimized with respect to all structural parameters, and the temperature dependence of coefficients $K_u$ and $K_{uv}$ defined in Eq.~(\ref{eq7}) is taken into account. Fig.~\ref{fig4} shows the comparison between the calculations obtained using our approach and the standard approach \cite{Bryce,Keblinski}, where the free energy is not minimized with respect to the internal atomic coordinate $\tau$ and elastic constants do not vary with temperature. Even though internal atomic position is relaxed as strain is applied in the standard method, this approach gives qualitatively very different trends compared to our model and experiments \cite{main,newmain}. These results highlight the importance of improving the standard method, to include the critical physical effects occurring near the phase transition, as shown here.

\begin{figure}[h!]
\begin{minipage}{0.49\textwidth}
\begin{center}
\includegraphics[width = 0.9\textwidth]{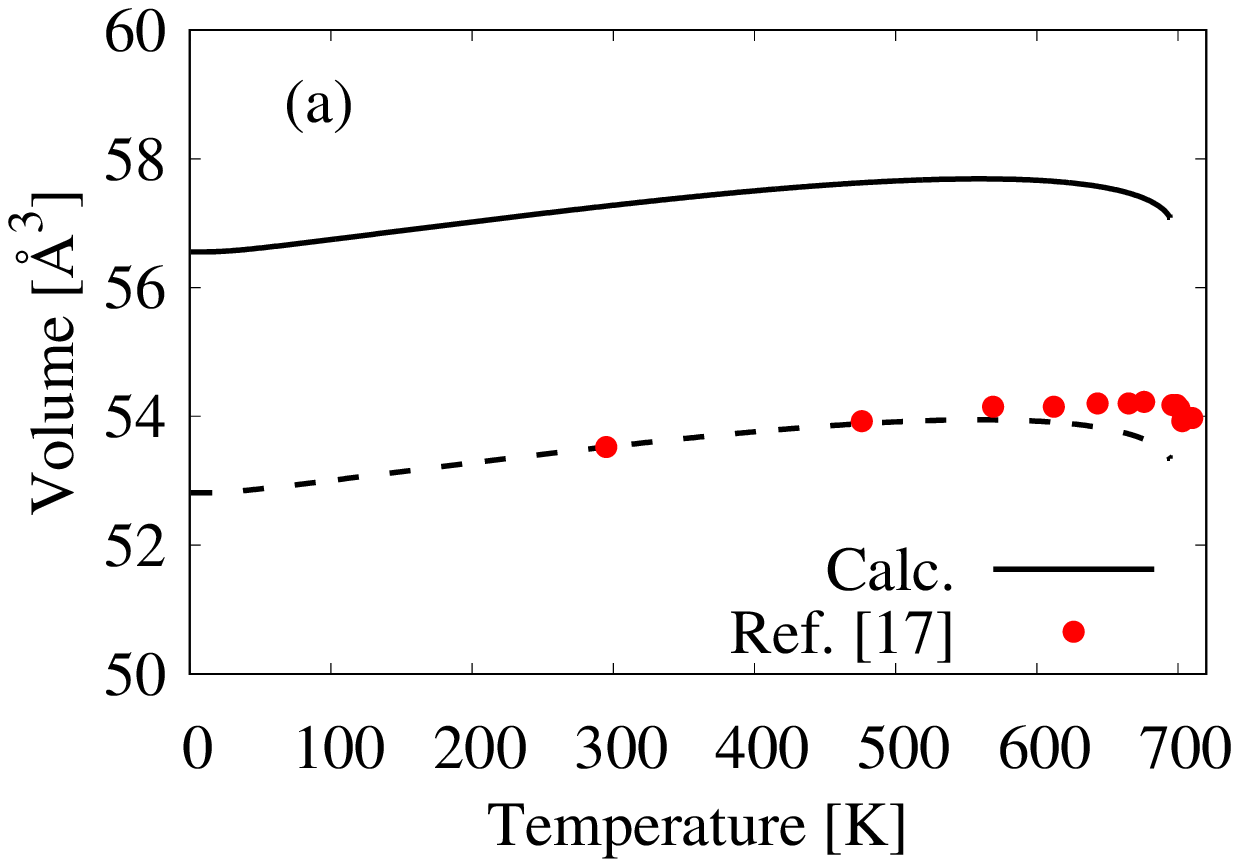}
\end{center}
\end{minipage}
\begin{minipage}{0.49\textwidth}
\begin{center}
\includegraphics[width = 0.9\textwidth]{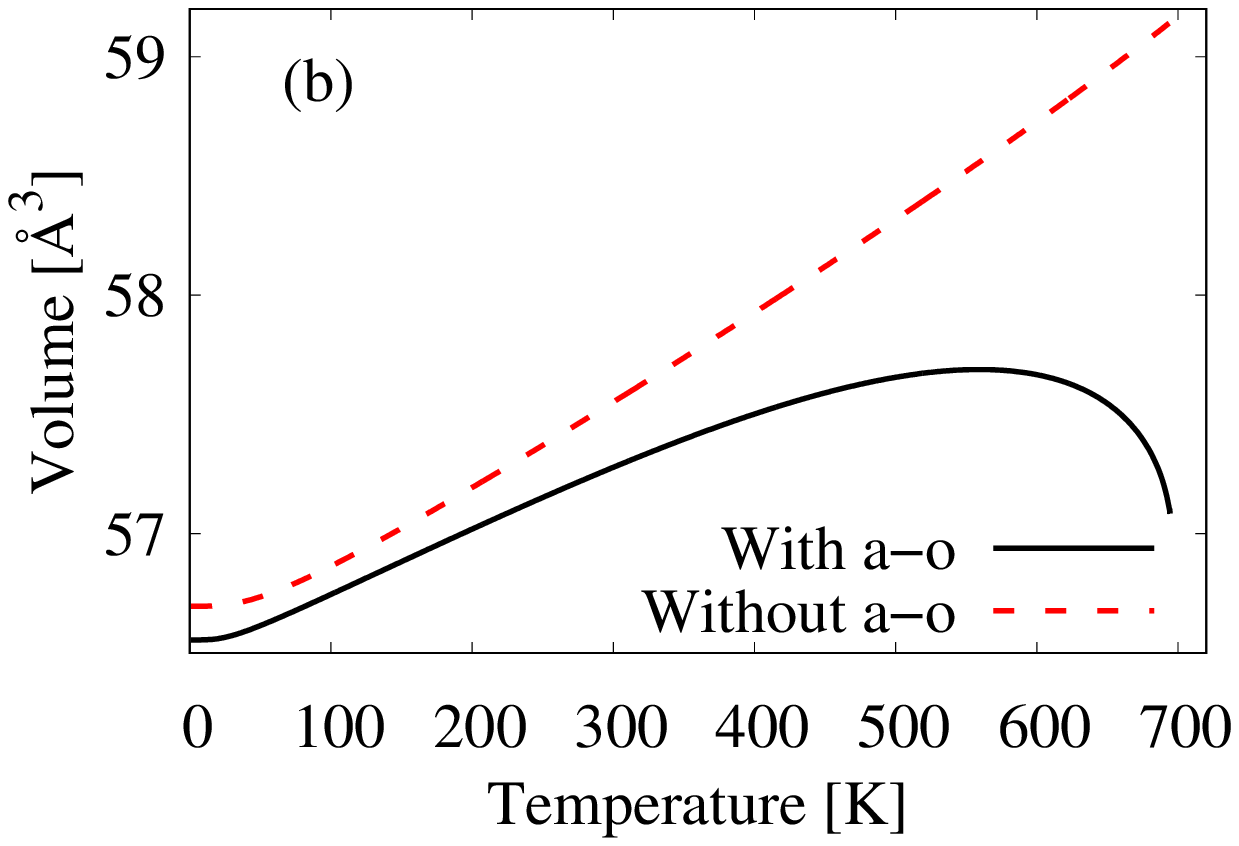}
\end{center}
\end{minipage}
\caption{(a) Volumetric thermal expansion of GeTe: our calculation (solid black line), experiment \cite{main} (red circles), and our calculation shifted by the difference between our and the experimental value at 300 K (dashed black line). (b) Computed volumetric thermal expansion including and neglecting acoustic-soft optical mode coupling, shown in solid black and dashed red lines, respectively.} 
\label{fig5}
\end{figure}

Most interestingly, GeTe exhibits negative volumetric thermal expansion near the phase transition at $\sim 700$ K, which has been observed experimentally \cite{main,newmain,abrikosov,Marchenkov1994} and reproduced in our calculations, see Fig.~\ref{fig5}(a). In contrast, the standard approach gives a positive volume expansion of GeTe in the whole temperature range considered. The volumetric contraction close to the phase transition is due to the NTE of the lattice constant shown in Fig.~\ref{fig3}(a). We note that the sign of the volumetric thermal expansion depends strongly on the exact composition of samples, as does the Curie temperature. Positive volumetric thermal expansion occurs in samples with more than $50.6$\% Te, as measured in Refs. \cite{Marchenkov1994, abrikosov}. Samples with less than $50.6$\% of Te exhibit NTE at the phase transition \cite{Marchenkov1994, abrikosov}, which is in agreement with our calculation for stoichiometric GeTe ($50\%$ Te). 

Analyzing all the physical quantities that determine the structural parameters (coefficients $K$ and generalized Gr{\"u}neisen parameters entering Eq.~(\ref{eq2})), we found that only $K_{a\tau}$, $K_{\theta\tau}$ and $K_{\tau\tau}$ change substantially near the phase transition.~(Elastic constants also vary considerably close to the transition, see Appendix C). $K_{a\tau}$ and $K_{\theta\tau}$ reflect static elastic energy variations with respect to simultaneous changes of the structural parameters related to acoustic strain ($a$ and $\theta$) and the TO mode ($\tau$). Consequently, $K_{a\tau}$ and $K_{\theta\tau}$ quantify acoustic-TO coupling, and indicate its large variation close to the phase transition.

Acoustic-soft TO mode coupling that increases considerably near the phase transition causes the negative thermal expansion of GeTe. In our computational method, we can artificially turn off this coupling by setting $K_{a\tau}$ and $K_{\theta\tau}$ to zero, as shown in Fig.~\ref{fig5}(b). The volume calculated by neglecting acoustic-TO coupling does not exhibit a negative thermal expansion. We thus conclude that strong acoustic-TO phonon coupling is the origin of the NTE of GeTe at the phase transition.

\begin{figure}[h!]
\begin{minipage}{0.49\textwidth}
\begin{center}
\includegraphics[width = 0.9\textwidth]{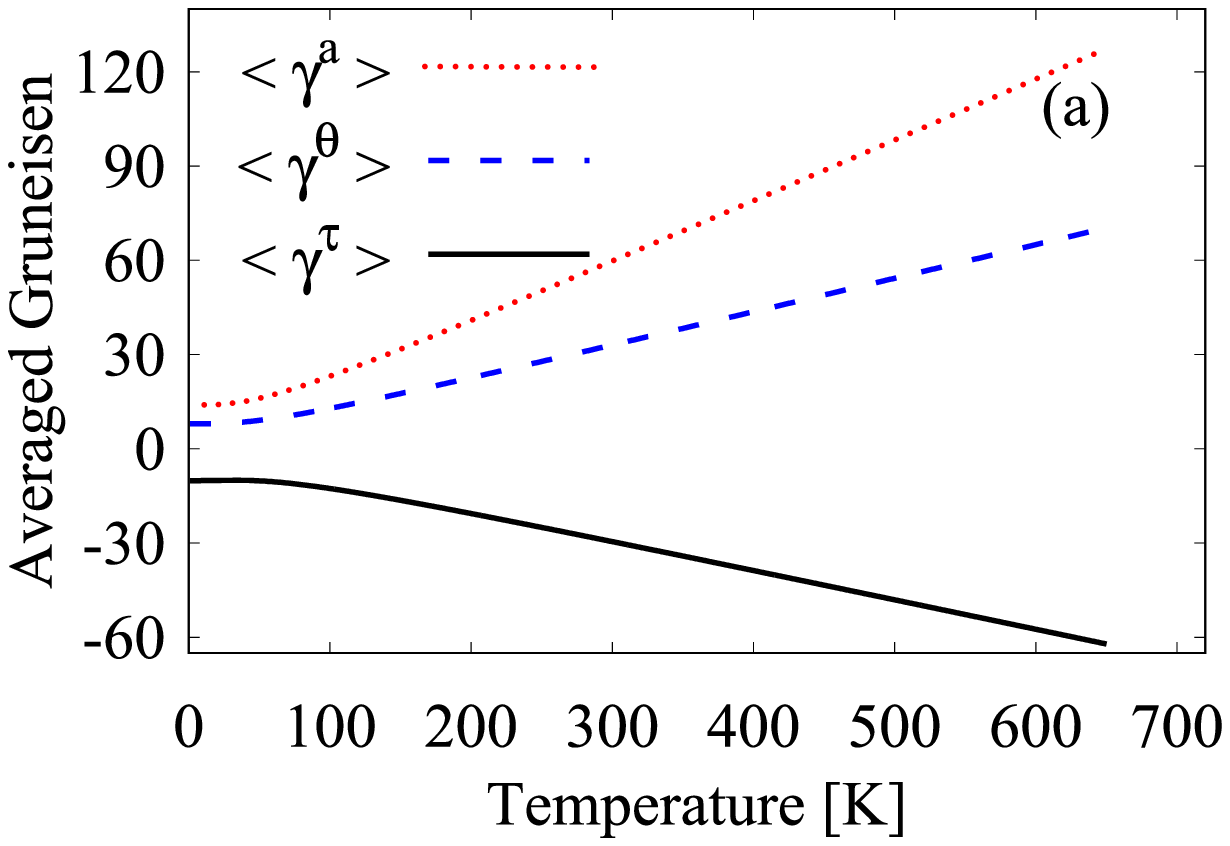}
\end{center}
\end{minipage}
\begin{minipage}{0.49\textwidth}
\begin{center}
\includegraphics[width = 0.9\textwidth]{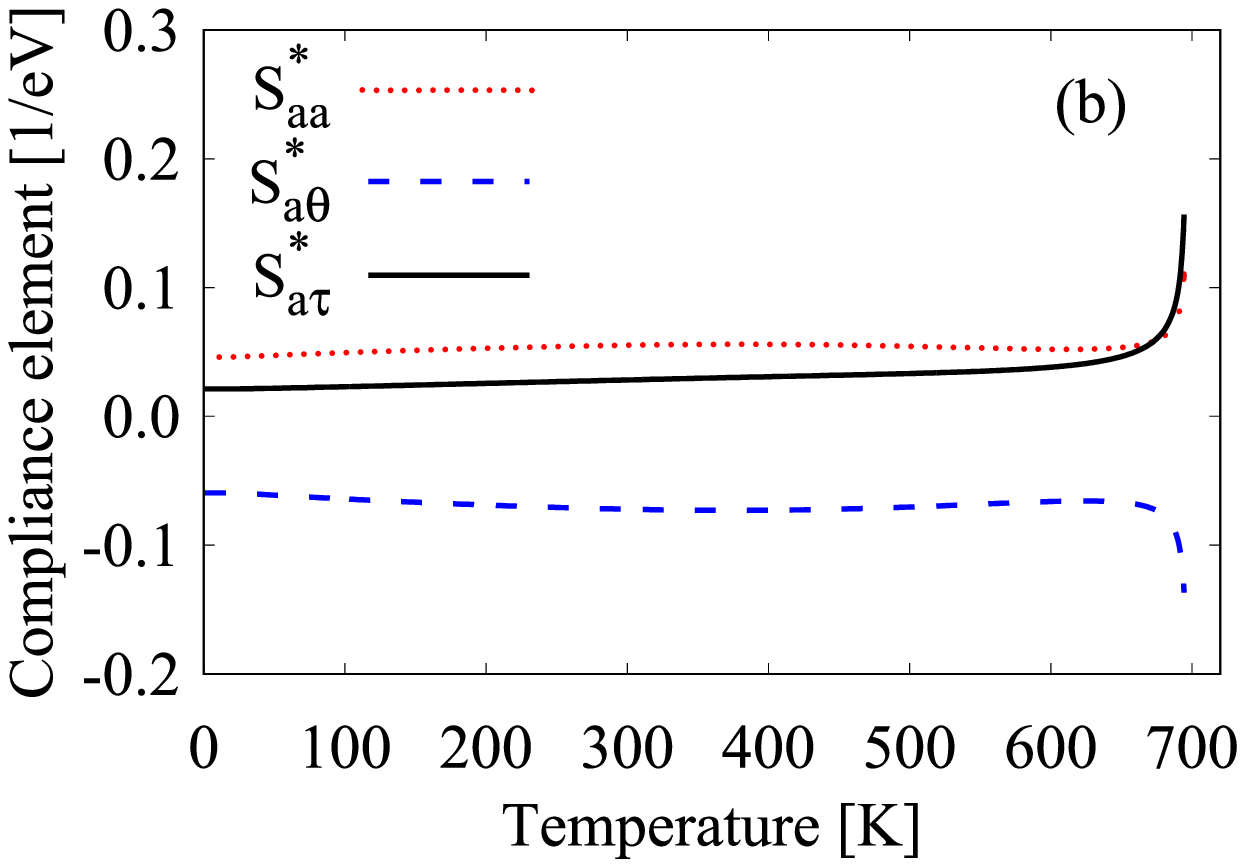}
\end{center}
\end{minipage}
\caption{Temperature dependence of: (a) average generalized Gr{\"u}neisen parameters defined for each structural parameter ($a$ - lattice constant, $\theta$ - angle, $\tau$ - internal atomic coordinate), and (b) normalized compliance matrix elements (see text for full explanation).}
\label{fig6}
\end{figure}

The most commonly cited cause of negative thermal expansion in the literature is a negative mode Gr{\"u}neisen parameter \cite{NTE1, NTE2, NTE3, NTE4}. Here we investigate the role of generalized Gr{\"u}neisen parameters in establishing the NTE of GeTe. We define average generalized Gr{\"u}neisen parameters for $u\in\{a,\theta,\tau\}$ as:
\begin{align*}
\left<\gamma ^{u}\right> =\frac{1}{\hbar \omega _{D}} \sum_{\textbf{q},s} \hbar\omega_{s}(\textbf{q})\left(n(\omega _{s}(\textbf{q})) + \frac{1}{2}\right)\gamma ^{u}_{s}(\textbf{q}), \numberthis
\label{eq17}
\end{align*}
where $\omega _{D}$ is the Debye frequency \cite{debye}. The temperature dependence of $\left<\gamma ^{u}\right>$ is shown in Fig.~\ref{fig6}(a). Fig.~\ref{fig6}(b) illustrates the compliance elements that determine the value of lattice constant in Eq.~(\ref{eq10}), normalized as $S_{aa}^*=S_{aa}/a^2$, $S_{a\tau}^*=S_{a\theta}/a\theta$, and $S_{a\tau}^*=S_{a\tau}/a\tau$. The linear temperature dependence of the average generalized Gr{\"u}neisen parameters stems from the Bose-Einstein occupation factor. In contrast, the compliance elements change dramatically with temperature near the phase transition, due to large temperature variations of $K_{a\tau}$, $K_{\theta\tau}$ and $K_{\tau\tau}$. Since the lattice constant expansion is proportional to $S_{aa}\left<\gamma ^{a}\right>+S_{a\theta}\left<\gamma ^{\theta}\right>+S_{a\tau}\left<\gamma ^{\tau}\right>$ (Eq.~(\ref{eq10})), its negative sign is partially due to negative $\left<\gamma ^{\tau}\right>$, which physically corresponds to the anharmonicity of the TO mode. Nevertheless, negative $\left<\gamma ^{\tau}\right>$ is not the main reason for NTE: it has to be accompanied by a large change of $S_{a\tau}$ i.e. large acoustic-TO coupling so that the expansion becomes negative. Furthermore, $S_{a\theta}$ is also negative and its absolute value increases more rapidly at the phase transition, resulting in an additional negative contribution to thermal expansion. This analysis confirms the dominant role of acoustic-TO coupling in establishing the NTE of GeTe near the phase transition. We expect that this conclusion will remain valid even when the temperature dependence of phonon frequencies and generalized Gr{\"u}neisen parameters $\gamma ^{u}_{s}(\textbf{q})$ is accounted for. This would make the temperature changes of $\left<\gamma ^{\tau}\right>$  near the phase transition somewhat larger than those calculated here, due to the temperature variations of the frequencies of soft TO modes close to the zone center. 

There is an ongoing debate in the literature about the true nature of the phase transition in GeTe (displacive vs order-disorder). Our method directly applies only to displacive phase transition. The experimental support for the displacive transition in GeTe was reported in Ref. \cite{main, newmain,displejsiv}. This is challenged by recent works of Fons \textit{et al.} \citep{Fons} and Matsunaga \textit{et al.} \cite{macunaga}, whose findings support the order-disorder picture. Our calculations show that the thermal expansion near the phase transition in GeTe can be well described with a purely displacive model. However, further investigation of order-disorder effects is needed for the complete description of the phase transition of GeTe.


\section{V. Conclusion}

We developed a first principles method that accurately describes the temperature dependence of all structural parameters for the rhombohedral phase of GeTe up to the Curie temperature of $\sim 700$ K. The key new features of our approach with respect to the standard method based on the Gr{\"u}neisen theory are the minimization of free energy with respect to all structural parameters, including internal atomic displacement, and the temperature dependence of static elastic energy. Our computed thermal expansion is in very good qualitative agreement with experiment. We showed that the coupling between acoustic and soft transverse optical modes is the main reason for the negative volumetric thermal expansion of GeTe near the phase transition. 

\begin{table*}[t]
\begin{center}
\begin{tabularx}{\textwidth}{ Y | c | Y | Y | Y | Y | Y }
\hline \hline 
  & $C_{11} + C_{12}$ [GPa] &$C_{13}$[GPa]&$C_{33}$[GPa]&$K_{aa}$[eV]&$K_{a\theta}$[eV] & $K_{\theta\theta}$[eV]\\
 \hline
  DFPT     & 114.756 & 29.962  & 63.899 & 72.764  & 74.514 & 27.243   \\ 
 \hline 
  Finite diff. DFT & 116.555 & 29.942  & 60.543 & 72.789    & 76.133 & 27.667    \\
\hline \hline
\end{tabularx}
\end{center}
\caption{Calculated elastic constants of GeTe using density functional perturbation theory (DFPT), and density functional theory (DFT) combined with a finite difference method. $C_{11} + C_{12}$, $C_{13}$ and $C_{33}$ were calculated directly using DFPT, and transformed into $K_{aa}$, $K_{a\theta}$ and $K_{\theta\theta}$ using Eq.~(\ref{eq15}). $K_{aa}$, $K_{a\theta}$ and $K_{\theta\theta}$ were computed using DFT, and transformed into $C_{11} + C_{12}$, $C_{13}$ and $C_{33}$ by inverting Eq.~(\ref{eq15}).} 
\label{tb2}
\end{table*}

\section{Acknowledgements}

This work was supported by Science Foundation Ireland (SFI) under Investigators Programme 15/1A/3160. We acknowledge the Irish Centre for High-End Computing (ICHEC) for the provision of computational facilities.

\section{Appendix A: Connection between our approach and standard approach to thermal expansion} 

Our approach for calculating the thermal expansion of rhombohedral materials can be linked to the standard method based on the Gr{\"u}neisen theory \cite{Bryce,Keblinski}. In contrast to our approach, the standard method minimizes the total free energy with respect to strain. Neglecting the contribution of internal atomic coordinate $\tau$, the elastic coefficients $K$ defined as the static elastic energy changes with respect to structural parameters in our method can be transformed into elastic constants. In the Voigt notation, the elastic matrix of a rhombohedral crystal reads:
\begin{equation}
\hat{C} =
\begin{bmatrix}
C_{11} & C_{12} & C_{13} & C_{14} & 0 & 0 \\
C_{12} & C_{11} & C_{13} & -C_{14} & 0 & 0 \\
C_{13} & C_{13} & C_{33} & 0 & 0 & 0 \\
C_{14} & -C_{14} & 0 & C_{44} & 0 & 0 \\
0 & 0 & 0 & 0 & C_{44} & C_{14} \\
0 & 0 & 0 & 0 & C_{14} & C_{66} 
\end{bmatrix},
\end{equation}
where $C_{66} = \frac{1}{2}(C_{11}-C_{12})$. Our coefficients $K$ can be converted to elastic constants via:
\begin{align*}
\label{eq15}
K_{aa} =& C_{11} + C_{12} + 2C_{13} + C_{33}/2,\\
K_{\theta\theta} =& Q_{1}^2 (C_{11} + C_{12}) -2Q_{1}Q_{2}C_{13} + Q_{2}^2 C_{33}/2, \numberthis \\
K_{a\theta} =& 2Q_{1}(C_{11}+C_{12}) + 2(Q_{1}-Q_{2})C_{13} -Q_{2}C_{33}.
\end{align*}
The coefficients $Q_{1}$ and $Q_{2}$ represent the dilatation of the hexagonal structure parameters (lattice constants perpendicular and parallel to the [111] axis) for unit dilatation of angle:
\begin{align*}
Q_{1} =& \frac{\sin \theta}{2(1-\cos \theta)}, \\
Q_{2} =& \frac{\sin \theta}{1+2\cos \theta}. \numberthis
\end{align*} 

We computed the elastic constant matrix $\hat{C}$ using density functional perturbation theory (\textsc{DFPT}) and \textsc{ABINIT} code, and transformed them into $K_{aa}$, $K_{a\theta}$ and $K_{\theta\theta}$ using Eq.~(\ref{eq15}). We also calculated $K_{aa}$, $K_{a\theta}$ and $K_{\theta\theta}$ using \textsc{DFT} and a finite difference method, and converted them into $\hat{C}$ by inverting Eq.~(\ref{eq15}). All elastic constants and coefficients $K$ obtained from \textsc{DFPT} and \textsc{DFT} calculations are in very good agreement, see Table \ref{tb2}. To the best of our knowledge, there are no reported experimental values for the elastic constants of GeTe. If we use the Voigt average for calculating the bulk modulus as:
\begin{align*}
9B = 2(C_{11} + C_{12}) + 4C_{13} + C_{33} = 2K_{aa}, \numberthis
\end{align*}
we obtain the value of $B = 45.92$ GPa at 0~K, which is in a good agreement with the experimental value of 49.9 GPa at 300~K \cite{bulk}.
 
We note that the elastic constants discussed above correspond to the clamped-ion elastic tensor $\hat{C}$, where the internal atomic coordinate is not relaxed in the presence of strain. We explicitly define the coefficients $K$ that take into account the relaxation of the internal atom: $K_{\tau\tau}$, $K_{a\tau}$, and $K_{\theta\tau}$. $K_{\tau\tau}$ represents the soft TO mode. $K_{a\tau}$ and $K_{\theta\tau}$ are related to the elements of the force-response internal-strain tensor as defined in Ref. \cite{intstrain}, and physically correspond to acoustic-soft optical mode coupling.

Now we identify the main differences between our approach and the standard approach to thermal expansion in the case of materials near phase transitions. Mode GP's in the standard approach are computed as:
\begin{align}
\frac{d \omega _{\lambda} (\mathbf{q})}{d \epsilon _{de}} = \frac{\partial \omega _{\lambda} (\mathbf{q})}{\partial \epsilon _{de}} + \frac{\partial \omega _{\lambda} (\mathbf{q})}{\partial \tau}\frac{\partial \tau}{\partial \epsilon _{de}} 
\label{eq19}
\end{align}
where $\epsilon _{de}$ is a component of the strain tensor~\cite{Bryce,Keblinski}. We consider a simplified expression for the total free energy of a rhombohedral system:
\begin{align}
F_{tot} =  K_{\tau\tau}\tau ^2 + K_{a\tau}a\tau + K_{\theta\tau}\theta\tau.
\end{align} 
To find the value of $\tau$ at thermal equilibrium, we minimize this function with respect to $\tau$:
\begin{align}
\frac{\partial F_{tot}}{\partial \tau} &= 2K_{\tau\tau}\tau + K_{a\tau}a + K_{\theta\tau}\theta = 0, \\
\tau &= -\frac{K_{a\tau}a + K_{\theta\tau}\theta}{2K_{\tau\tau}}.
\end{align}
We estimate the terms that correspond to the term $\partial \tau/\partial \epsilon _{de}$ in Eq.~\eqref{eq19} by replacing $\epsilon _{de}$ with $a$ (or $\theta$):
\begin{align}
\frac{\partial \tau}{\partial a} = -\frac{K_{a\tau}}{2K_{\tau\tau}}.
\end{align} 
The coefficient $K_{\tau\tau}$ corresponds to the zone center soft TO mode, and becomes zero at the phase transition. Our calculations show that $K_{a\tau}$ is finite at the phase transition. Consequently, the factor $\partial \tau/\partial a$ diverges at the phase transition. Our method captures the temperature dependence of elastic coefficients $K_{uv}$, $u,v\in \{a,\theta,\tau\}$, and thus the temperature dependence of the terms $\partial \tau/\partial a$ and $\partial \tau/\partial \theta$. In contrast, the standard method gives the corresponding terms only at 0~K. Both methods ignore the temperature dependence of the terms $\partial \omega _{\lambda} (\mathbf{q})/\partial \epsilon _{de}$ and  $\partial \omega _{\lambda} (\mathbf{q})/\partial \tau$ in Eq.~\eqref{eq19}. We stress that the temperature dependence of elastic coefficients is critical for the description of the NTE of GeTe near the phase transition. This can be obtained straightforwardly by explicitly accounting for $\tau$ in the free energy minimization, as done in our method.

\section{Appendix B: Soft optical mode frequency} 

\begin{figure}[h]
\begin{center}
\includegraphics[width=0.441\textwidth]{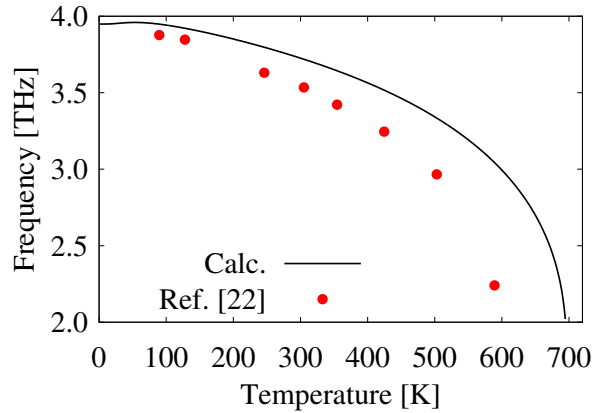}
\caption{TO mode frequency versus temperature: our calculation (solid black line) and experiment \cite{STEIGMEIER19701275} (red circles).}
\label{fig8}
\end{center}
\end{figure}

Since $K_{\tau\tau}$ is the second derivative of total energy with respect to internal atomic coordinate, we can calculate the TO mode frequency using \cite{srivastava}:
\begin{align}
\omega _{TO} ^2 = \frac{2K_{\tau\tau}}{\mu a_{||}^2}, 
\end{align}
where $\mu$ is reduced mass of the unit cell and $a_{||}$ is the length of the unit cell in the [111] direction. The temperature dependent elastic coefficient  $K_{\tau\tau}$ is computed as:
\begin{align}
K_{\tau\tau} &= K^{0} _{\tau\tau} + 3K^{0} _{\tau\tau\tau}\delta \tau + K^{0} _{\tau\tau\theta}\delta \theta + K^{0} _{\tau\tau a} \delta a\nonumber\\ & + 6K^{0} _{\tau\tau\tau\tau}\delta \tau ^2  + 3(K^{0} _{\tau\tau\tau\theta}\delta \theta + K^{0} _{\tau\tau\tau a} \delta a)\delta \tau  \\
 &+ K^{0} _{\tau\tau\theta\theta}\delta \theta ^2 + K^{0} _{\tau\tau aa}\delta a ^2 + K^{0} _{\tau\tau\theta a}\delta \theta \delta a.\nonumber
\end{align}
Consequently, the anharmonic contribution to the zone center TO mode frequency is explicitly accounted for in our model up to the second order. The coefficients $K^{0} _{\tau\tau\tau}$ and $K^{0} _{\tau\tau\tau\tau}$ describe the anharmonicity of the soft TO mode energy potential. The coefficients such as $K^{0} _{\tau\tau a}$, $K^{0}_{\tau\tau\theta}$, $K^{0} _{\tau\tau\tau a}$ etc. describe anharmonic acoustic-soft TO mode coupling. We also account for the temperature dependence of $a_{||}$. As result, we can track the softening of TO mode as a function of temperature and compare it to measurements~\cite{STEIGMEIER19701275}, as shown in Fig.~\ref{fig8}. We find a very good agreement between our calculated TO frequency and experiment. 

\section{Appendix C: Elastic constants near the phase transition} 

\begin{figure}[h!]
\begin{minipage}{0.49\textwidth}
\begin{center}
\includegraphics[width = 0.9\textwidth]{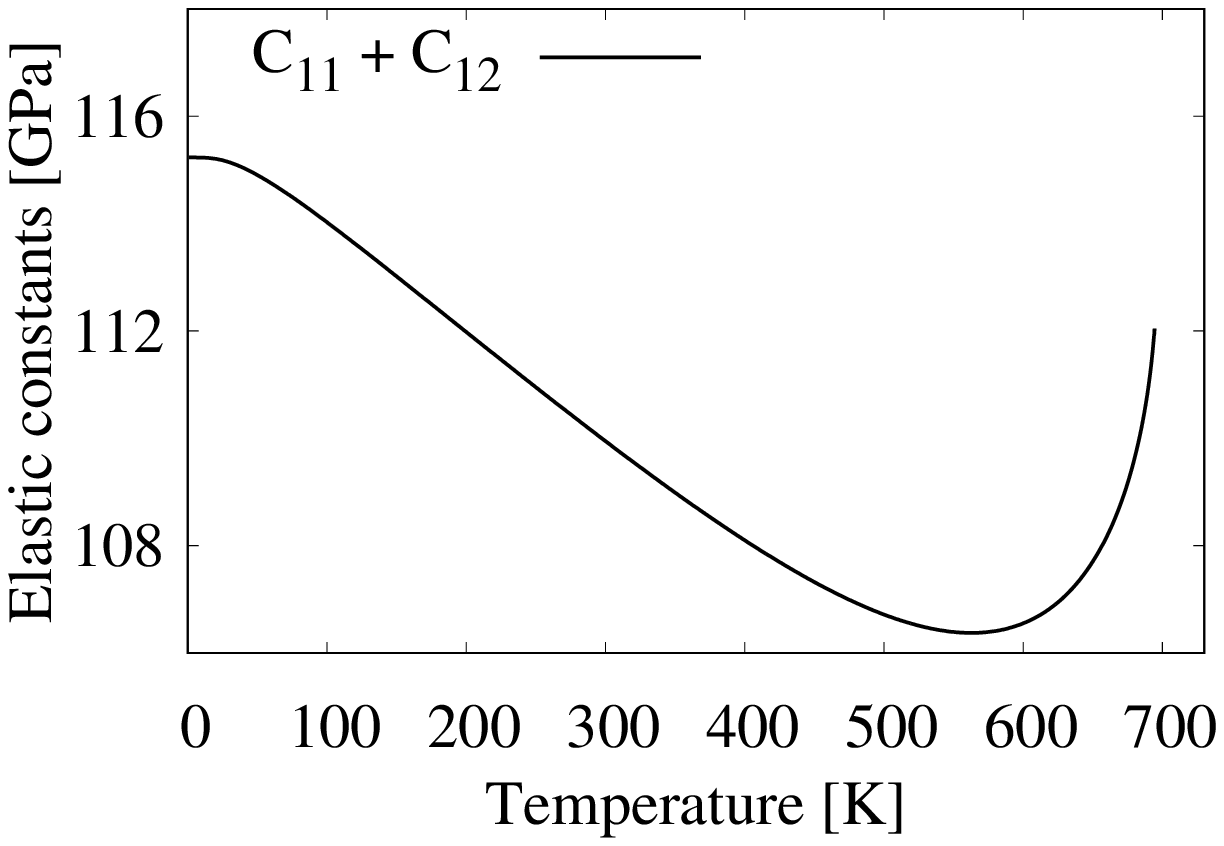}
\end{center}
\end{minipage}
\begin{minipage}{0.49\textwidth}
\begin{center}
\includegraphics[width = 0.9\textwidth]{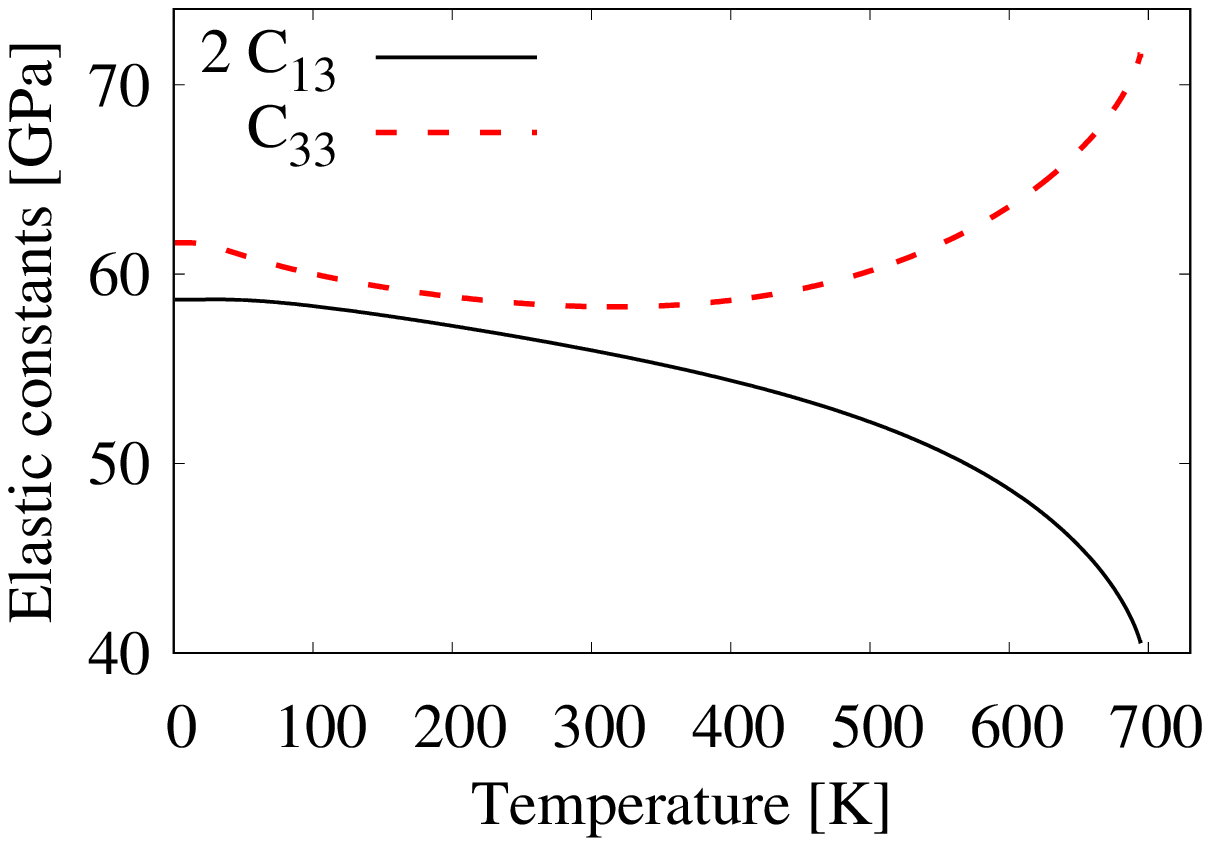}
\end{center}
\end{minipage}
\caption{Elastic constants of GeTe as functions of temperature: (a) $C_{11} + C_{12}$, (b) $C_{13}$ (solid black line) and $C_{33}$ (dashed red line).}
\label{fig7}
\end{figure}

In our calculations, the values of all elastic constants have a steep change at the phase transition, which is in agreement with experimental observations in Sn$_{x}$Ge$_{1-x}$Te \cite{elastSn} and Pb$_{x}$Ge$_{1-x}$Te \cite{elastPb}. Fig. \ref{fig7} shows how $C_{11} + C_{12}$, $C_{13}$ and $C_{33}$ vary with temperature. $C_{11} + C_{12}$ increases rapidly at the phase transition, as observed in \cite{elastPb, elastSn}. Experimental values of $C_{13}$ and $C_{33}$ were not reported, but our calculations correctly capture their expected behaviour. At the high symmetry rocksalt phase, $C_{33}$ and $C_{11}$, as well as $C_{12}$ and $C_{13}$, should have the same values. In the low symmetry rhombohedral phase, $C_{33}$ has lower value than $C_{11}$ (Table \ref{tb1}), so we would expect that $C_{33}$ will increase towards the phase transition to become equal to $C_{11}$. On the other hand, $C_{13}$ is larger than $C_{12}$, and it will decrease towards the phase transition to become equal to $C_{12}$. Both of these trends are observed in our results.  

We made an attempt to verify whether our calculated values of elastic constants satisfy the Born criteria for mechanical stability:
\begin{align*}
&C_{11} - C_{12} > 0, \\
&C_{44} > 0, \numberthis \\
&C_{11} + 2C_{12} > 0.
\end{align*}
In our calculations, which are restricted to rhombohedral symmetry structures, we cannot separately calculate the elastic constants $C_{11}$ and $C_{12}$, and can only track their sum. Our DFPT calculation at 0~K gives $C_{11} \gg C_{12}$ ($C_{11} = 93.33$ GPa, $C_{12}$ = 21.43 GPa). Since $C_{11}+C_{12}$  does not vary substantially with temperature, see Fig.~\ref{fig7}(a), it is likely that $C_{11}$ and $C_{12}$ individually exhibit a similar trend. This suggests that the relations $C_{11} \gg C_{12}>0$, $C_{11} - C_{12} > 0$ and $C_{11} + 2C_{12} > 0$ should remain valid up to the Curie temperature. We cannot track the elastic coefficient $C_{44}$ related to shear strain since we do not allow symmetry lowering types of strain.

\bibliography{GeTeThermalExp}

\begin{thebibliography}{66}
\bibitem [{deb()}]{debye}%
  \BibitemOpen
  \href@noop {} {}\bibinfo {howpublished} {We calculated the Debye frequency of
  GeTe by fitting the analytical expression for the specific heat capacity
  obtained using the Debye model to that calculated using the phonon dispersion
  computed using density functional theory. Our calculated value of the Debye
  frequency is $3.8$ THz.}\BibitemShut {Stop}%
\end{thebibliography}

\end{document}